\documentclass[%
 aip,
 jmp,
 amsmath,amssymb,
preprint,%
]{revtex4-1}

\usepackage{graphicx}
\usepackage{dcolumn}
\usepackage{bm}
\usepackage[mathlines]{lineno}

\usepackage[utf8]{inputenc}
\usepackage[T1]{fontenc}
\usepackage{mathptmx}
\usepackage{etoolbox}
\usepackage{blindtext}
\usepackage[colorlinks=true,linkcolor=black,citecolor=black,urlcolor=black]{hyperref}
\makeatother
\begin{document}
\preprint{AIP/123-QED}

\title{Flat-top electron velocity distributions driven by wave-particle resonant interactions}

\author{S. Zanelli}
\email{sofia.zanelli@unical.it}
\affiliation{Dipartimento di Fisica, Universitá della Calabria, Rende, I-87036, Italy}
\author{S. Perri}
\affiliation{Dipartimento di Fisica, Universitá della Calabria, Rende, I-87036, Italy}
\author{M. Condoluci}
\affiliation{Dipartimento di Fisica, Universitá della Calabria, Rende, I-87036, Italy}
\author{P. Veltri}
\affiliation{Dipartimento di Fisica, Universitá della Calabria, Rende, I-87036, Italy}
\author{F. Pegoraro}
\affiliation{Dipartimento  di Fisica “Enrico  Fermi”, Università di Pisa, Pisa, 56122, Italy }
\affiliation{Istituto di Astrofisica e Planetologia Spaziali, via Fosso del Cavaliere 100, Roma, 00133, Italy}
\author{O. Pezzi}
\affiliation{Istituto per la Scienza e Tecnologia dei Plasmi, Consiglio Nazionale delle Ricerche via Amendola 122/D, I-70126, Bari, Italy}
\author{D. Perrone}
\affiliation{ASI- Italian Space Agency, Via del Politecnico SNC, I-00133 Rome, Italy}
\author{D. Trotta}
\affiliation{European Space Agency (ESA), European Space Astronomy Centre (ESAC), Camino Bajo del Castillo s/n, E-28692 Villanueva de la Canada, Madrid, Spain}
\author{F. Valentini\footnote{Corresponding author: francesco.valentini@unical.it}}
\affiliation{Dipartimento di Fisica, Universitá della Calabria, Rende, I-87036, Italy}

\date{\today}

\begin{abstract}
\textbf{ABSTRACT}\\
The role of kinetic electrons in the excitation and sustainment of ion-bulk electrostatic waves in collisionless plasmas is investigated, with a focus on the physical mechanisms responsible for the generation of small-scale structures in space plasmas. Building on the work of F. Valentini et al., PRL, 106, 165002 (2011), we numerically solve the Vlasov-Poisson system in one spatial and one velocity dimension for both ions and electrons. Our findings reveal that a significant fraction of the energy supplied by an external driving electric field, used to trigger ion-bulk waves excitation, is transferred to electrons, which become trapped within the wave potential well. As a result, multiple phase-space vortices, generated during the early time evolution, undergo a merging process in the long-time limit, ultimately resulting in a single, coherent, and persistent phase-space hole in the distributions of both species. Furthermore, the resonant interaction between electrons and ion-bulk fluctuations induces a velocity-space diffusion process, leading to the development of a "flat-top" profile in the electron velocity distribution, routinely observed in near Earth space. To establish observational relevance, virtual spacecraft measurements were performed to evaluate the detectability of the velocity distribution features observed in the simulations using modern spaceborne instruments. The results presented here are consistent with observations of electrostatic phenomena in space plasmas, and underscore the widespread occurrence of such structures across various plasma environments.
\end{abstract}

\maketitle

\section{\label{sec:level1} Introduction}
The study of collisionless plasma dynamics at small scales is vital for understanding the mechanisms of energy transfer, dissipation, and structure formation in space plasmas. In such environments, where collisional effects are negligible and energy cascades from macroscopic to microscopic scales, fundamental questions about the processes responsible for particle heating and the emergence of small-scale structures arise, which are stiff matter of debate \citep{valentini2011new, bruno2013solar}. Observational studies \citep{gurnett1977plasma, gurnett1978ion} have revealed a wealth of electrostatic wave activity in space plasmas, with measurements capturing features such as ion-acoustic waves and other high-frequency phenomena. Recent advancements, such as those by \citet{graham2016electrostatic, graham2021kinetic, graham2015electrostatic}, have further documented the ubiquitous presence of small-scale electrostatic activity in the solar wind and in the Earth's magnetosphere, offering new insights into their formation and dynamics.

A pivotal contribution to this field was the identification of ion-bulk (IBk) waves by \citet{valentini2011new}, which represent a novel branch of electrostatic modes driven by particle trapping processes \citep{o1965collisionless, valentini2006excitation, holloway1991undamped}. 
These waves exhibit acoustic-type dispersion with phase speed near the ion thermal velocity and can persist under conditions that would heavily damp traditional ion-acoustic waves, that is, for instance, values of the electron to ion temperature ratio of order unity. Indeed, as demonstrated in Ref.~\onlinecite{valentini2011new}, a successful excitation of these fluctuations requires an external driver electric field applied to the plasma, to trap resonant ions and flatten the ion velocity distribution in the vicinity of the phase speed of the external field, this finally inhibiting Landau damping \citep{landau196561}.
This makes IBk waves particularly relevant in diverse space plasma settings where the electron-to-ion temperature ratio, $T_e/T_i$, is typically close to unity. Several combined observational and numerical studies (see, for example, \cite{valentinietal2014,vecchioetal2014,perri2021nature}) contributed to the identification of these IBk fluctuations in the solar wind and in the terrestrial magnetosheath. 
As an example, in 2021, \citet{perri2021nature} identified IBk fluctuations in regions characterized by intense electrostatic activity  in the terrestrial magnetosheath, through the analysis of observation measurements from the NASA Magnetospheric MultiScale (MMS) space mission. In these regions, the ion velocity distribution function (VDF) displays a pronounced plateau (shoulder/beam) along the direction of the local magnetic field, in the vicinity of the ion thermal speed.
However, previous numerical analyses employed a linear fluid Boltzmann approximation for electrons, thereby neglecting potential kinetic effects that could significantly influence wave-particle interactions and the resultant plasma dynamics.

In this work, we extend the investigation of IBk waves by solving the Vlasov-Poisson system numerically in one spatial and one velocity dimension for both ions and electrons, employing a realistic value of the ion-to-electron mass ratio of $m_{i}/m_{e}=1836$. This choice, while computationally expensive, ensures a more accurate and realistic representation of plasma dynamics \citep{ng2024kinetic}. In our numerical experiments, an external driving electric field is employed to enable the trapping of resonant protons in the wave potential well, in the velocity range around the wave’s phase velocity. This process effectively suppresses Landau damping \citep{landau196561} and leads to the excitation of IBk fluctuations. This external driving electric field plays a role analogous to the resonant interactions between protons and circularly polarized ion-cyclotron waves propagating parallel to the background magnetic field, which is routinely at work in real space plasma systems \citep{valentini2008cross, valentini2009electrostatic, valentini2011short, bowen2022situ, trotta2024properties}. In fact, according to the Kennel-Engelmann theory \citep{kennel1966velocity}, such interactions can produce diffusive plateaus in the longitudinal proton velocity distribution by redistributing particle energy through wave-particle resonance \citep{klein2020diagnosing, valentini2005self}.
{Moreover, we note that the idea of external driving has several applications in heliospheric plasmas. Indeed, flat-top distributions are found downstream of Earth’s bow shock, strong source of quasi-static electrostatic potential~\citep[see][]{lalti2024adiabatic}. Future efforts include the study of flat-top distributions at interplanetary shocks where electrostatic wave activity is also enhanced \citep{boldu2024ion}. This effect may also be relevant to magnetic reconnection events, where flat-top distributions are routinely observed as well~\citep{hoshino2001strong, asano2008electron}.

In contrast to previous studies that employed the fluid electron approximation \citep{valentini2011new}, where the energy supplied by the external driver was entirely dedicated to trapping resonant protons and thereby exciting the IBk branch, the present simulations reveal that a significant fraction of the driver energy is also utilized for electron trapping. Our numerical results demonstrate that: (i) the excitation of the IBk branch remains achievable, even though part of the driver energy is redirected to electron trapping; (ii) IBk fluctuations can persist against electron-driven Landau damping when the realistic electron-to-proton mass ratio is considered and, over extended timescales, the system sustains persistent electrostatic activity, consistent with recent spacecraft observations \citep{graham2016electrostatic, graham2021kinetic, graham2015electrostatic}; (iii) after the external driver is switched off and the IBk branch is successfully excited, the resonant interaction between electrons and IBk fluctuations initiates a velocity-space diffusion process \citep{krall1974principles}, resulting in the formation of a flat region near $v \simeq 0$ in the electron velocity distribution — a characteristic "flat-top" profile commonly observed in spacecraft measurements \citep{graham2021non, asano2008electron, masood2022formation, zhao2017electron, qureshi2019whistler, stasiewicz2024origin, aravindakshan2023theory}. 
Our numerical results provide an alternative perspective on the origin of electron flat-top velocity distributions, complementing previous studies such as, for instance, (i) Ref.~\onlinecite{stasiewicz2024origin}, which focused on the Earth’s bow shock region and attributed the formation of these distributions to a stochastic wave energization mechanism, (ii) Ref.~\onlinecite{zhao2017electron}, where the development of flat-top electron distributions was associated with current sheet dynamics producing magnetic reconnection in the geo-magnetic tail and resulting in omnidirectional flat-top profiles, and (iii) Ref.~\onlinecite{dum1974turbulent}, in which, by means of numerical simulations, the generation of the flat-top electron velocity distributions was associated with the quenching of the current-driven ion-sound instability \citep{sagdeev1969nonlinear}.
By incorporating kinetic electron dynamics, our study demonstrates that in a realistic electrostatic plasma, where both protons and electrons are treated as kinetic species (with a realistic mass ratio), the excitation of IBk fluctuations is not only viable but also leads to the emergence of flat-top velocity profiles in the electron distribution as a direct consequence of the kinetic electron response to the propagation of these electrostatic fluctuations.
The paper is organized as follows. In Section~\ref{sec:numericalmodel}, we describe the numerical model and the simulation setup, detailing the methods employed to solve the Vlasov-Poisson system and the parameters used in our analysis. Section~\ref{sec:results} presents the numerical results, focusing on the excitation and sustainment of ion-bulk waves and the corresponding phase-space dynamics of protons and electrons. In Section~\ref{mimicking}, we discuss the application of virtual spacecraft measurements to assess the detectability of the velocity distribution features observed in our simulations using modern spaceborne instruments. Finally, in Section~\ref{sec:summary}, we summarize our findings and outline potential directions for future research.

\section{Numerical model and simulation set up}\label{sec:numericalmodel}
In this section, we describe the numerical algorithm used to solve the Vlasov-Poisson equation system, treating both ions and electrons as kinetic species to reproduce the excitation of nonlinear IBk waves. We consider ions with an atomic number $Z=1$, corresponding to protons. Throughout this work, time is normalized by the inverse proton plasma frequency $\omega_{p,p}^{-1}$, velocities are scaled by the proton thermal speed $v_{th,p}$, and lengths are scaled by the proton Debye length $\lambda_{D,p}$. 

Under these normalizations, the Vlasov-Poisson system in a 1D-1V phase-space configuration is written as:
\begin{eqnarray}
& &\frac{\partial f_{p}}{\partial t} + v\frac{\partial f_{p}}{\partial x} + \left(E+E_{D}\right)\frac{\partial f_{p}}{\partial v} = 0, \label{eqn:fp} \\
& &\frac{\partial f_{e}}{\partial t} + v\frac{\partial f_{e}}{\partial x} - \frac{m_{p}}{m_{e}}\left(E+E_{D}\right)\frac{\partial f_{e}}{\partial v} = 0, \label{eqn:fe} \\
& &\frac{\partial E}{\partial x} = \left[\int_{-\infty}^{+\infty} f_{p} dv - \int_{-\infty}^{+\infty} f_{e} dv \right], \label{eqn:E}
\end{eqnarray}
where $f_{p}=f_{p}(x,v,t)$ and $f_{e}=f_{e}(x,v,t)$ are the distribution functions for protons and electrons, respectively. Moreover, $E=E(x,t)$ represents the self-consistent electric field, while $E_{D}=E_{D}(x,t)$ denotes the external electric field used to trap resonant protons and electrons. The analytical form for $E_D$ is given below.

The simulation phase space is discretized with $N_{x}=512$ grid points in the spatial domain, imposing periodic boundary conditions. In the velocity domain, we use $N_{k,e}=2001$ grid points for electrons and $N_{k,p}=1401$ for protons. The velocity limits are defined as $v_{max,p} = 5$ for protons and $v_{max,e} = 6v_{th,e}$ for electrons, where $v_{th,e} = [(T_e/T_p)(m_p/m_e)]^{1/2}$ is the normalized electron thermal speed. The simulation parameters include a temperature ratio of $T_{e}/T_{p} = 5$ and a mass ratio of $m_{p}/m_{e} = 1836$. The spatial domain length is $L = 2\pi \times 20 = 125.664$, corresponding to a wave number $k = 2\pi / L=0.05$. 

At equilibrium, both species have uniform density $n_{0} = 1$ and Maxwellian velocity distributions. This equilibrium is perturbed by the external driver $E_D$, with the form:
\begin{equation}
E_{D}(x,t) = E_{D}^{max} \left\{ 1 + \left[ (t - \tau) / \Delta \tau \right]^{2n} \right\}^{-1}\sin{(kx-
\omega_D t)}
\end{equation}
where $E_{D}^{max} = 0.05$, $\tau = 2600$, $\Delta \tau = 2000$, $n = 15$, and $\omega_{D}=kv_{\phi,D}$, with $v_{\phi,D} = 2.5$ which represents the driver's phase speed. 
This value of phase velocity of the driver was selected based on the results of \citet{valentini2011new}, which demonstrated that, for a temperature ratio of $T_e/T_p = 5$, this value maximizes the plasma response. This ensures optimal conditions for the excitation of IBk waves in our simulations.
This driver is applied to both protons and electrons, and its amplitude decays to near-zero at approximately $t \approx 5000$, with a total simulation duration of $t_{max}=32500$. 
It is crucial to point out that any abrupt turn on or off of the external driver field would excite Langmuir waves and ion-acoustic waves together with ion-bulk waves, thus  complicating the analysis. Thus, the driver is turned on and off adiabatically.

The Vlasov equation is solved using the time-splitting scheme introduced by \citet{cheng1976integration} (see also \citet{pezzi2013eulerian, celebre2023phase, pezzi2016collisional} for further details). For evolution in both physical and velocity space, we implemented an upwind finite-volume scheme based on the work by \citet{van1997towards}, achieving third-order accuracy in $\Delta x$ and $\Delta v_\alpha$, with errors proportional to $\Delta x^4$ and $\Delta v_\alpha^4$. The Poisson equation is solved using a Fast Fourier Transform (FFT) algorithm \citep{cooley1965algorithm}, ensuring efficient and accurate computation of the electric field. The time step $\Delta t$ is chosen to satisfy the well-known Courant-Friedrichs-Lewy stability condition, ensuring numerical stability throughout the simulation \citep[CFL][]{courant1967partial}. The total energy is monitored during the entire simulation, with energy variations remaining consistently below $10^{-4}\%$.

\section{Numerical results}\label{sec:results}
Building on previous results \citep{valentini2011new, valentini2011excitation}, we investigate the excitation of IBk waves, explicitly incorporating the kinetic response of electrons. To derive the theoretical dispersion relation for the IBk waves, one needs to solve for the zeros of the electrostatic dielectric function, $D\left(k,\omega\right)$ \citep{krall1974principles}, which accounts for contributions from both protons and electrons. Assuming the presence of a plateau of vanishing velocity width in the velocity distribution of both species near the wave phase speed, the imaginary part of the dielectric function vanishes (undamped solutions). Consequently, we numerically search for the roots of the real part $D_{R}(k,\omega_{R})$ of the dielectric function. 
As a result, the solution for the real part of the wave frequency $\omega_{R}$ as a function of the wave number $k$ (that is the curve where the roots of $D_{R}(k,\omega_{R})$ are located) is shown in black in Fig. ~\ref{fig:reldisp} for $T_{e}/T_{p}=5$.

The dispersion relation, shown in Fig.~\ref{fig:reldisp} and commonly known as the \textquotedblleft tear-drop curve\textquotedblright, reveals two distinct acoustic branches: the lower branch corresponds to the IBk waves, while the upper branch represents the standard ion-acoustic (IA) waves. The red line depicts the theoretical prediction for the real part of the frequency of ion-acoustic (IA) waves, as described in \citet{krall1974principles}. The yellow line indicates the curve $\omega_R=1.7 k$ (which, in the range of small wavenumbers, is tangent to the tear-drop curve for the IBk branch), and the vertical red-dashed line marks the wave number $k = 0.05$. It is important to note that these predictions are derived in ideal conditions where the plateaus in the velocity distributions have vanishing width. For finite velocity width plateaus, the phase speed of the IBk waves shifts to higher values, as discussed in \citet{valentini2006excitation, valentini2011excitation, valentini2011new}.

Using the two-species Vlasov-Poisson code described in section \ref{sec:numericalmodel}, we numerically reproduce the excitation of IBk waves, specifically focusing on the kinetic dynamics of protons and electrons. In Fig.~\ref{fig:Figura2}(a), the time evolution of the electric field at a fixed spatial position $x_0 = L/2$ is shown in black, while the time evolution of the driver amplitude is represented by the red line. It can be observed that, once the driver is turned off, the oscillation amplitude persists over time, and the electric field oscillates at an almost constant saturation value.
In panel (b) of the same figure, a zoomed-in view of the electric field $E(x_0, t)$ is displayed, corresponding to the region delimited by the  vertical red-dashed lines in panel (a), specifically within the time interval $t \in [12000, 15000]$. It is worth noting that the Fourier mode $m=1$ has been filtered out from the electric signal in panel (b), as this mode is directly driven by the external forcing and remains dominant over the other modes, even after the driver is switched off. By filtering out mode $m=1$, the observed signal represents the plasma's intrinsic response to the external driver. From this plot, it is evident that the time behavior is erratic, characterized by numerous spikes and a superposition of several Fourier components (retaining the contribution of mode $m=1$ would have partially masked the erratic nature of the fluctuations).
To better emphasize the  point, the left panel of Fig.~\ref{fig:electricfield_amplitude_and_spettro} shows the time evolution of the absolute value of the amplitude of the first 10 electric field Fourier components with $m = 1,\cdots , 10$. It is evident that all wave numbers are excited during the driving process and persist even after the external driver is turned off. The spectral component $m = 1$, directly driven by the external electric field, is depicted in black and emerges as the dominant component by the end of the simulation. However, the energy content of the other Fourier components remains significant.
To identify the nature of the nonlinear electrostatic fluctuations excited by the external driver, we compute the $k-\omega_R$ spectrum of the numerical electric field signal, which provides the dispersion relation of the fluctuations. This spectrum is displayed in the right panel of Fig.~\ref{fig:electricfield_amplitude_and_spettro}, where a clear and well-defined acoustic branch is recovered. The black-dashed line in this panel represents the curve $\omega_R = 2.5k$, indicating that the electrostatic fluctuations triggered by the external driver consist of multiple wave numbers, each propagating with the same phase velocity $v_\phi \simeq 2.5$. This acoustic-like branch of waves can be identified as the IBk waves, according to \citet{valentini2011excitation}. As demonstrated here, IBk waves can be excited even in the presence of kinetic electrons, a scenario previously ruled out in earlier studies \citep{valentini2011excitation, valentini2011new}.
The spectrum of the electric energy is shown in the log-log plot of Fig.~\ref{fig:spettro}, where the dependence of the spectral electric energy $|E_k|^2$ on the wave number $k$ is presented at four distinct times, marked by the vertical dashed lines in the left panel of Fig.~\ref{fig:electricfield_amplitude_and_spettro}. At later times, the spectrum spans approximately two decades of wave numbers with a nearly constant slope, indicating that energy flows towards higher wave numbers during the system evolution, resembling a turbulent cascade.
At this point, to understand how the driver energy is distributed among protons, electrons, and the self-consistent electric field, we analyzed the time evolution of the electric energy and of the kinetic energies of protons and electrons, defined as follows:

\begin{eqnarray}
& & E_{kin,p} = \frac{1}{2} \int_0^{L_x} \mathrm{d}x \int_{-\infty}^{\infty} v^2 f_p \, \mathrm{d}v \\
& & E_{kin,e} = \frac{m_e}{2 m_p} \int_0^{L_x} \mathrm{d}x \int_{-\infty}^{\infty} v^2 f_e \, \mathrm{d}v \\
& & E_{el} = \int_0^{L_x} \frac{E^2}{2} \, \mathrm{d}x
\end{eqnarray}

At each time step, these quantities were numerically evaluated, with the velocity integration performed over the intervals $[-v_{max,p}, v_{max,p}]$ and $[-v_{max,e}, v_{max,e}]$, respectively.
In Fig.~\ref{fig:energy}, the black curve represents the variation $\Delta E_{kin,p}(t) = E_{kin,p}(t) - E_{kin,p}(0)$ for protons, the red curve indicates the same variation for electrons, while the blue curve represents the variation of the electric energy. As can be clearly seen from this plot, the largest fraction of the energy injected by the driver is transferred to the protons. However, a significant fraction of this energy is also transferred to the electrons. It is worth noting that the electron contribution to the total energy, represented by the red curve in Fig.~\ref{fig:energy}, was excluded in earlier works \citep{valentini2011excitation, valentini2011new}.

In the process of resonant wave-particle energy exchange, the driver energy is used to trap both resonant protons and electrons in the wave potential well, ultimately leading to the generation of multiple phase space vortices, centered around the wave phase speed, in the particle distribution functions.
This is illustrated in Fig.~\ref{fig:contour}, where the phase space contour plots of the proton (top row) and electron (bottom row) distribution function are shown at four different times (from left to right), throughout the simulation. 
During the driving process [panels (a) and (b)], multiple vortices are generated, corresponding to different Fourier components of the electric field (see Fig.~\ref{fig:electricfield_amplitude_and_spettro}). Over time, these structures exhibit a tendency to merge and collapse \citep{berk1970phase, ghizzo1988stability, manfredi2000stability, depackh1962water, valentini2008decay, affolter2018trapped, dubin2018parametric} into a single phase space structure in the long-term limit [panel (d)].

As previously discussed, these vortical structures are generated near the phase velocity of the fluctuations ($v_\phi = 2.5$). Owing to the substantial difference in thermal velocities between protons and electrons—attributable to their mass disparity—the formation of these structures occurs over distinct velocity ranges. For protons, the vortical structures arise at velocities exceeding the thermal speed, whereas for electrons, they form closer to $v = 0$, as shown in Fig.~\ref{fig:contour}. It should be noted that this phenomenology would be significantly affected by an artificially reduced mass ratio, often adopted for computational convenience. Furthermore, the trapping region for electrons is considerably broader than that for protons when expressed in terms of proton thermal velocities, highlighting a larger phase-space confinement for electrons.

In Fig.~\ref{fig:fde_all}, we present the one-dimensional electron velocity profiles at different times during the simulation. These profiles are obtained by averaging the electron distribution functions shown in the bottom row of Fig.~\ref{fig:contour} over the spatial range of $\simeq 25 \lambda_{D,p}$, defined by the vertical red-dashed lines in each corresponding plot. For clarity and ease of comparison, each velocity profile is normalized to its maximum value. As evident from this plot, during the external driving phase (black curve), the electron velocity distribution is perturbed as electrons begin to become trapped in the wave’s potential well. Over time, after the external driver is switched off, a velocity diffusion process ensues due to the resonant interaction between trapped electrons and IBk waves. This leads to an expansion of the wave-particle interaction region and a progressive flattening of the velocity profile, as indicated by the red and yellow curves. In the long-time limit, an extensive flat region, approximately $60 v_{th,p}$ in width and centered near $v \simeq 0$, becomes clearly visible (blue curve). This feature is commonly referred to as a flat-top velocity distribution and is frequently observed in spacecraft measurements of space plasma environments \citep{graham2021non, asano2008electron, masood2022formation, zhao2017electron, qureshi2019whistler, stasiewicz2024origin, aravindakshan2023theory}.
Figure~\ref{fig:fd_vs_v} provides a comparison between the velocity profiles of the proton (left panel) and electron (right panel) distribution functions, spatially averaged over the intervals marked by the vertical red-dashed lines in Fig.~\ref{fig:contour}(d). The proton distribution exhibits a significantly perturbed region, manifesting as a shoulder structure \citep{araneda2008proton, valentini2008cross} centered around $v \simeq v_\phi$. 
The generation of this distortion is a typical signature of the resonant interaction of particles with fluctuations, driven by a velocity diffusion process which, in the framework of the quasi-linear theory \citep{krall1974principles}, leads to the generation of a flat velocity region (plateau) in the vicinity of the wave phase speed.
As shown previously by the blue curve in Fig.~\ref{fig:fde_all}, a flat-top velocity profile emerges for electrons, with a velocity width comparable to the electron thermal speed, as indicated by the vertical red-dashed lines.
Our findings provide a physical explanation for the formation of these flat-top velocity distributions through a velocity diffusion process driven by the resonant interaction of IBk fluctuations and trapped electrons. This process can be efficient even at low value of the electron to proton temperature ratio, where standard ion-acoustic waves are heavily Landau damped. This interaction redistributes electron velocities, resulting in the characteristic flattened region centered at $v = 0$, which matches those observed in space plasma measurements. This mechanism underscores the significant role of wave-particle interactions in shaping the electron velocity distributions seen in natural plasma environments.

\section{Mimicking real spacecraft measurements of the particle velocity distributions} \label{mimicking}
The purpose of the following analysis is to determine whether the modifications and perturbations in the proton and electron distribution functions observed in the above simulations can be detected using modern instruments onboard spacecraft. To this end, we employed the virtual instrument technique, which involves running synthetic measurements within the simulation to mimic real observations in space. As the measuring capabilities of modern missions increases together with the quality of numerical simulations, virtual spacecraft become a unique tool, capable of one-to-one comparisons between theory and observations. Thus, the technique developed here is transversal and will be applied in future studies to different cases characterized by the emergence of complex velocity distributions, such as the region upstream of interplanetary shocks~\citep[e.g.,][]{Preisser2020,Lario2022}.

Measurements of particle velocity distribution functions in space can be performed using top-hat electrostatic analyzers \citep{carlson1982instrument, paschmann1998analysis}. For instance, the Fast Plasma Investigation (FPI) on the Magnetospheric Multiscale (MMS) mission measures the differential directional flux of electrons and ions with unprecedented temporal resolution, enabling the study of kinetic-scale plasma dynamics. Each of the four MMS spacecraft is equipped with eight top-hat spectrometers for electrons and eight for ions, strategically placed to achieve full angular coverage of the field of view. The instruments for electrons and ions are distinct and consist of two concentric hemispheres with an aperture on the outer hemisphere, and the hemispheres are set at different voltages \citep{pollock2016fast}. The electric field between the hemispheres allows particles with a specific energy-per-charge ratio $\left(E/q\right)$ to pass through the aperture and reach the detector. By varying the applied voltage, particles are sorted based on their $E/q$. Particles entering the analyzer parallel to its axis are focused onto specific sectors of the detector, with each sector corresponding to a distinct azimuthal velocity direction. Top-hat analyzers typically have a 360-degree disk-shaped field of view, and to sample the full $4\pi$ solid angle, either spacecraft rotation or electrostatic polar deflectors are employed. 

The energy-angular resolution of the top-hat analyzer is crucial for studying kinetic-scale processes since inadequate phase-space resolution would blur the fine structures in the particle distribution functions. Thus, the top-hat instrument counts charged particles and categorizes them by charge, energy, and direction of arrival, producing a histogram of these counts on a three-dimensional energy-angular polar grid.

To model the virtual response of such an instrument within our numerical simulations, we developed an algorithm to generate a three-dimensional histogram of particle counts on a polar grid, emulating a real top-hat instrument. For a standard top-hat analyzer, the polar grid is uniformly spaced in polar angle $\left(0 \leq \theta \leq \pi\right)$ and azimuthal angle $\left(0 \leq \phi \leq 2\pi\right)$, and logarithmically spaced in energy. The grid spacing is defined as follows:

\begin{equation}
    \theta_i = i\Delta\theta; \quad i = 0, \ldots, N_\theta; \quad \Delta\theta = \pi/N_\theta
\end{equation}
\begin{equation}
    \phi_j = (j-1)\Delta\phi; \quad j = 1, \ldots, N_\phi; \quad \Delta\phi = 2\pi/N_\phi
\end{equation}
\begin{equation}
    E_k=E_{min}\left (\frac{E_{max}}{E_{min}}\right)^{k/N_E}; \quad k=0,\cdots, N_E
\end{equation}

We generated $5 \times 10^{5}$ protons and $5 \times 10^{5}$ electrons, distributing their velocities along the $x$-direction based on the one-dimensional velocity profiles shown in Fig. \ref{fig:fd_vs_v}. Velocities in the $y$- and $z$-directions followed Maxwellian distributions, consistent with the densities and temperatures of protons and electrons used in the simulation. Naturally, in real scenarios, three-dimensional velocity distributions can also be shaped and distorted in the $y$- and $z$-directions by various effects—such as particle interactions with other types of fluctuations—which are not included in the model discussed in this paper. As a result, these distributions may deviate from the typical Maxwellian configuration even along $y$- and $z$-directions. Our primary focus here, however, is to determine whether the perturbations generated exclusively along one direction ($x$-direction), under the electrostatic approximation, are detectable by spacecraft instruments.

We incorporated the actual parameters of the top-hat electrostatic analyzer onboard NASA's Magnetospheric Multiscale (MMS) mission \citep{burch2016magnetospheric, pollock2016fast}, which explores the near-Earth space environment (solar wind, magnetosphere). Table \ref{tab:MMS} summarizes the instrument's key parameters and the typical mean speeds of protons and electrons observed in the magnetospheric environment. 

\begin{table}[ht]
\centering
\small
\begin{tabular}{|c|c|c|c|c|c|c|}
\hline
 & $N_\theta$ & $N_\phi$ & $N_E$ & $E_{min}$ (eV) & $E_{max}$ (eV)  & $V$ (km/s) \\ \hline
$p$   &    18      &    34    &  33   & 9.38                & 28258.65            &  157.41    \\ \hline
$e$ &    18      &    34    &  33   & 9.68                & 27586.20            &  151.55    \\ \hline
\end{tabular}
\caption{Typical parameters used for mimicking the virtual top-hat electrostatic analyzer,. based on the instrument onboard the NASA MMS mission for protons and electrons.}
\label{tab:MMS}
\end{table}

Using these parameters, we distributed the simulated protons and electrons onto an energy-angular grid, mimicking the data output of a real instrument, and generated three-dimensional histograms. The results are shown in Fig. \ref{fig:Top_hat}, where the $\left(E, \phi\right)$ plane at $\theta = \pi/2$ is presented for protons (left) and electrons (right). For protons, the core of the velocity distribution and the particle beam in the positive $x$-direction are clearly resolved \citep{marsch1985beam, perri2021nature, perri2020deviation}. For electrons, a central white hole corresponds to the minimum energy threshold of the instrument; nonetheless, the surrounding blue region represents the flat-top structure observed in one dimension in the simulation.

This analysis demonstrates that the velocity distributions generated in the numerical simulation, arising from the excitation and propagation of IBk waves, can indeed be detected with the energy-angular resolution typical of the top-hat instrument onboard NASA's MMS mission.

\section{Summary and Conclusions}\label{sec:summary}

In this study, we explored the kinetic role of electrons in the excitation and sustainment of ion-bulk (IBk) electrostatic waves in collisionless plasmas, providing new insights into the small-scale processes that govern energy transfer in space environments. Building upon prior work that utilized a Boltzmann approximation for electrons, we numerically solved the Vlasov-Poisson system in one spatial and one velocity dimension for both ions and electrons. This approach allowed us to uncover critical aspects of wave-particle interactions and emphasized the role of IBk waves in shaping plasma dynamics at kinetic scales. Importantly, we adopted a realistic mass ratio of $m_{p}/m_{e} = 1836$, which, as highlighted earlier, plays a crucial role in the generation of the flat-top profile in the electron velocity distribution. 
Indeed, if a realistic proton-to-electron mass ratio is employed, the IBk phase speed (of the order of $v_{th,p}$) is much lower than the electron thermal velocity, thus the electron resonant region (the velocity region where electrons interact resonantly with IBk fluctuations) falls close to $v\simeq 0$. On the other hand, when the mass ratio is artificially decreased, due to, for example, computation reasons, the electron resonant region moves to larger velocities correspondingly, departing from $v\simeq 0$, where it is detected in the space data.

Our findings demonstrate that the external driver effectively excites the IBk wave branch, even when the kinetic response of electrons is accounted for. Spectral analysis of the numerical signals confirms the presence of the acoustic branch associated with IBk waves, characterized by multiple wave numbers propagating at the same phase velocity. These fluctuations persist even after the external driver is switched off, underscoring the capability of IBk waves to efficiently channel energy to smaller scales along a turbulent cascade.

The propagation of IBk fluctuations is accompanied by rich phase-space dynamics. While protons absorb the majority of the energy from the external driver, a substantial fraction is transferred to electrons, leading to the formation of trapped populations and multiple phase-space vortices for both species. Over time, these phase-space vortices undergo a merging process, with smaller structures collapsing into larger ones, eventually stabilizing into a configuration characterized by a single, coherent phase-space hole for both protons and electrons. In the long-term evolution, the resonant interaction of trapped electrons with IBk fluctuations drives a velocity-space diffusion process that produces flat-top velocity distributions, a feature commonly observed in space plasmas. These long-lived structures persist over extended timescales, demonstrating the potential of IBk waves to sustain nonlinear electrostatic activity in collisionless plasma environments.
To assess the observational relevance of our results, we employed a virtual instrument approach to evaluate whether the velocity distribution features obtained in our simulations for protons and electrons could be detected using current instrumentation. The significance of these synthetic measurements is twofold: first, they provide a powerful tool to interpret observations from current and past spacecraft missions in controlled environments, as demonstrated by the analysis of MMS measurements in this study. Second, virtual spacecraft techniques are essential for the design of future missions, enabling the testing of observational capabilities and benchmarking instrument performance prior to launch, as in the case of the M-class candidate mission Plasma Observatory. By simulating the response of top-hat electrostatic analyzers, such as those onboard NASA's MMS mission, we verified that these instruments have sufficient resolution to detect the flat-top electron profiles and proton beam structures generated in our simulations. This confirms the detectability of such features in space plasmas and highlights the significance of IBk wave dynamics as a fundamental mechanism for interpreting observed electrostatic phenomena in natural plasma environments.
Future research will aim to extend the results discussed in the present paper to multi-dimensional configurations, incorporating additional factors such as magnetic fields and multi-species interactions, by means, for instance, of fully-kinetic Eulerian Vlasov-Maxwell models \citep{pezzi2019vida}. Such studies will further elucidate the interplay of wave-particle dynamics and the role of IBk waves in shaping space plasma behavior at kinetic scales.

\begin{acknowledgments}
SP and FV acknowledge discussions with Julia E Stawarz. This project has received funding from the European Union’s Horizon Europe research and innovation programme under grant agreement No. 101082633 (ASAP) and from the ASI project “Attività di Fase A per la missione Plasma Observatory” (2024-15-HH.0).
FV, OP and DP acknowledge the support of the PRIN 2022 project “The ULtimate fate of TuRbulence from space to laboratory plAsmas (ULTRA)” (2022KL38BK, Master CUP: B53D23004850006), funded by the Italian Ministry of University and Research. The numerical simulations presented in this paper were performed on the NEWTON cluster at the University of Calabria. This study was carried out within the Space It Up project funded by the Italian Space Agency, ASI, and the Ministry of University and Research, MUR, under contract n. 2024-5-E.0 - CUP n. I53D24000060005. SP acknowledges the project `Data-based predictions of solar energetic particle arrival to the Earth: ensuring space data and technology integrity from hazardous solar activity events' (CUP H53D23011020001) funded by the Italian Ministry of University and Research.
\end{acknowledgments}

\nocite{*}
\bibliography{biblio_ZANELLI}
\bibliographystyle{apsrev4-1}

\newpage


\begin{figure}[ht]
\begin{center}
\includegraphics[width=0.50\textwidth]{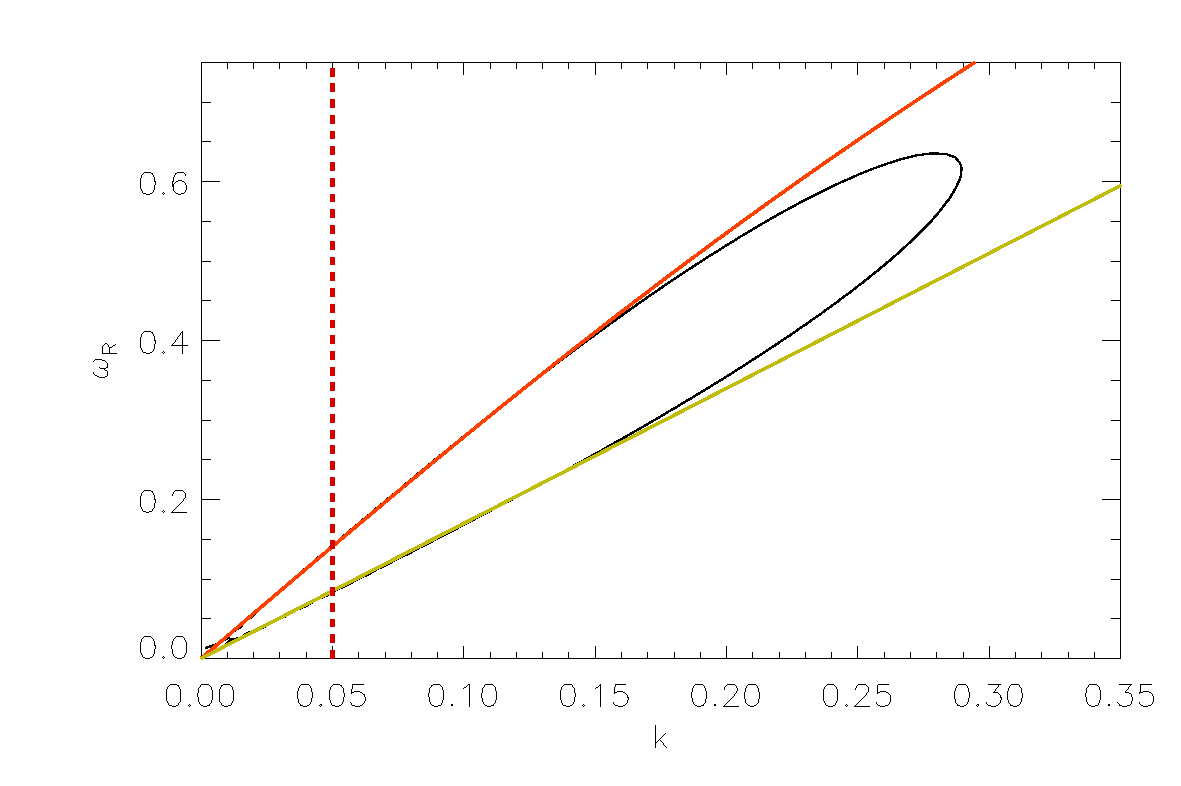}
\end{center}
\caption{Real part of the frequency as a function of the wave number for \( T_{e}/T_{p} = 5 \) (black curve). The red solid line represents the theoretical dispersion relation for ion-acoustic waves, while the yellow line indicates the curve $\omega_R=1.7 k$ in the low-wavenumber regime. The vertical red-dashed line denotes \( k = 0.05 \).}

\label{fig:reldisp}
\end{figure}

\begin{figure}[ht]
    \centering
    \hspace{-0.8cm}
    \includegraphics[width=0.52\textwidth]{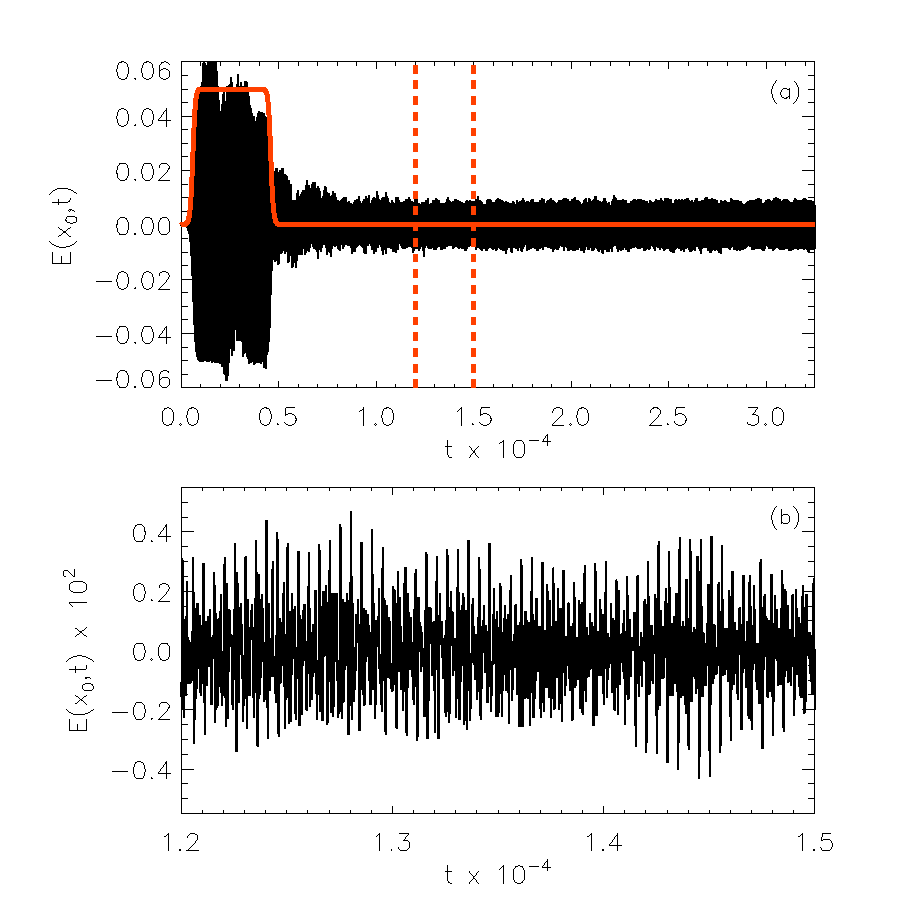} 
    \caption{Panel (a): Time evolution of the electric field at a fixed spatial position \( x_{0} \). The red curve represents the amplitude of the external driver. Panel (b): Zoom of the electric field signal after removing the contribution from the Fourier mode \( m=1 \), within the time interval \( t \in [12000, 15000] \) (indicated by the two vertical red-dashed lines in the top panel).}
    \label{fig:Figura2}
\end{figure}

\begin{figure*}[ht]
    \centering
    \includegraphics[width=0.49\textwidth]{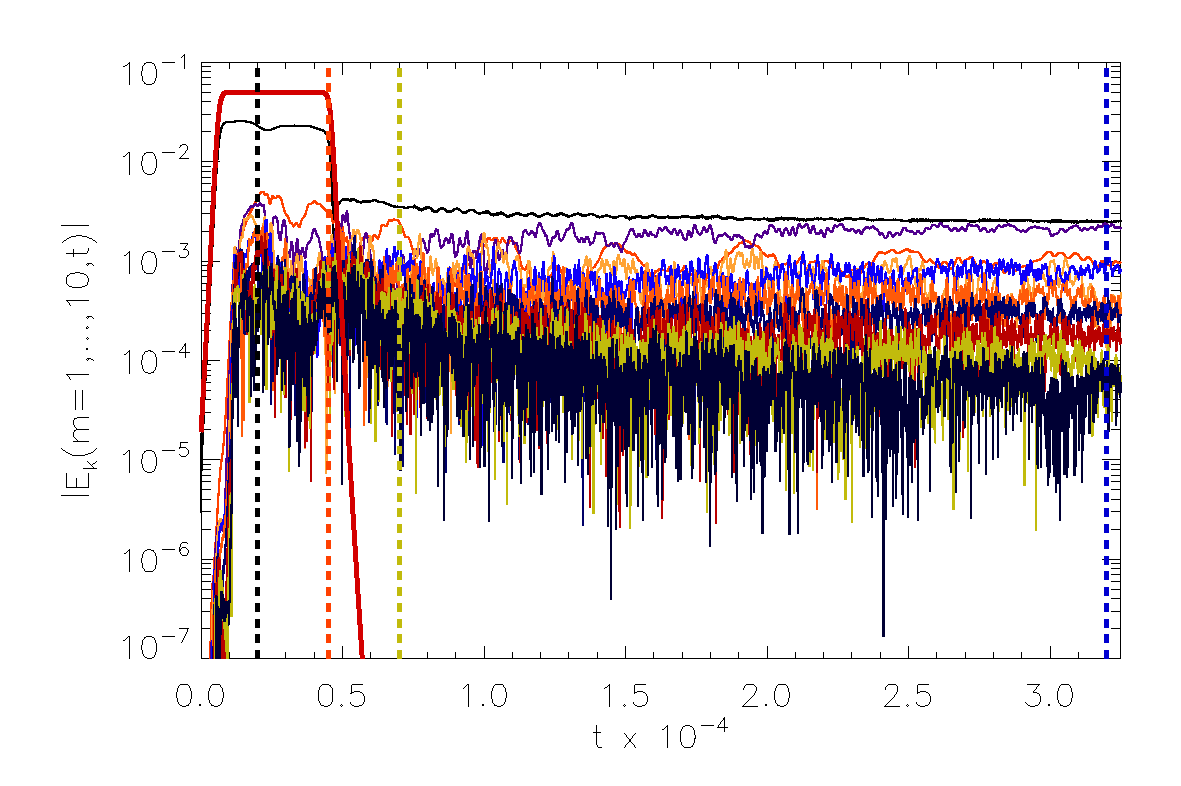} \hfill
    \includegraphics[width=0.49\textwidth]{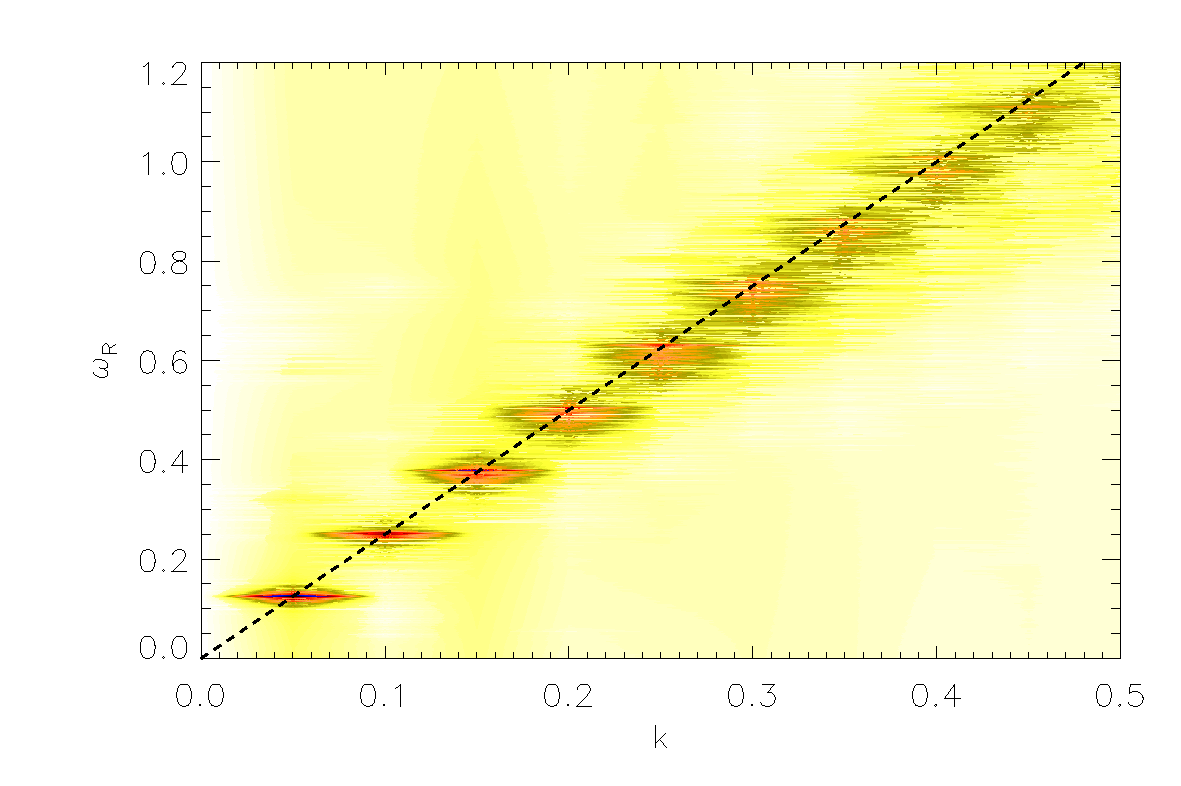}
  \caption{Left plot: Time evolution of the electric field amplitude for the first 10 Fourier modes, shown on a semi-logarithmic scale. The \( m=1 \) mode is represented in black, while the amplitude of the external driver is depicted in dark red. Vertical dashed lines indicate the time instances corresponding to the spectral electric energy shown in Fig. \ref{fig:spettro}. Right plot: \( k-\omega_{R} \) spectrum of the numerical electric field signal. The black-dashed line represents the curve \(\omega_{R} = 2.5k\).}
    \label{fig:electricfield_amplitude_and_spettro}
\end{figure*}

\begin{figure}[ht]
    \centering
    \includegraphics[width=0.49\textwidth]{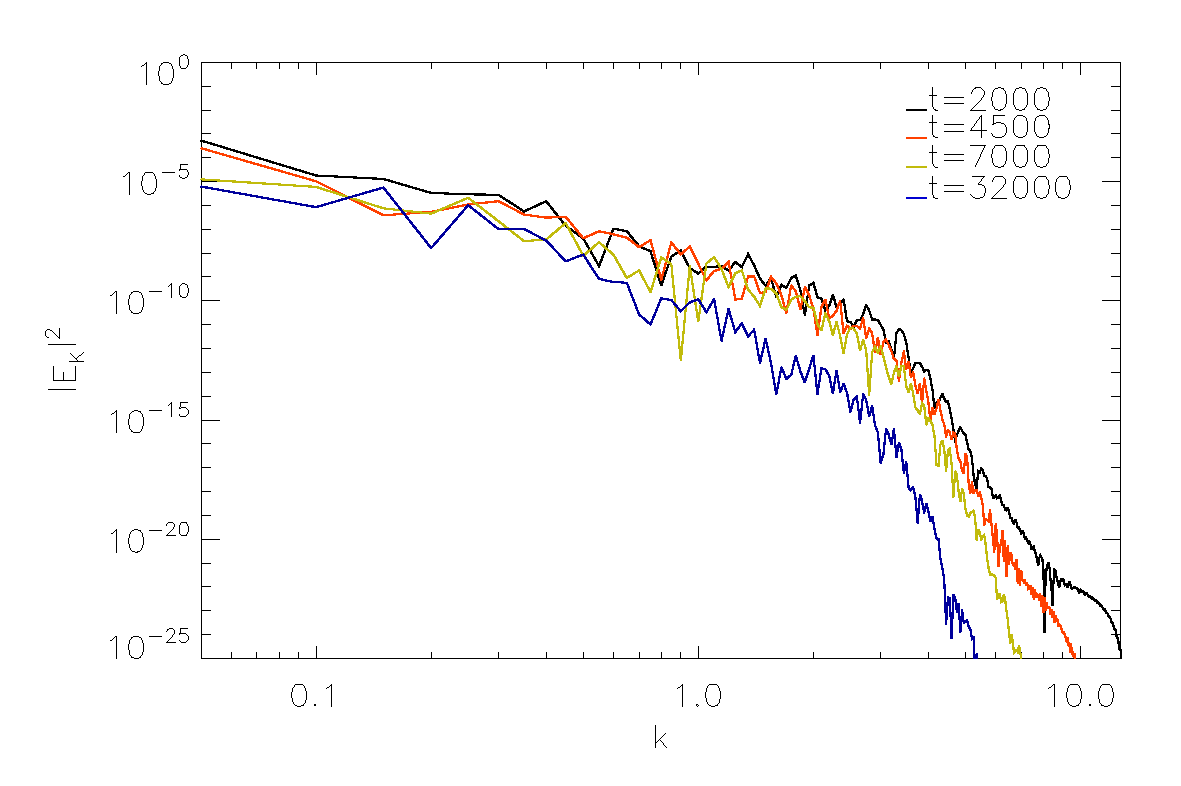}
  \caption{Log-log plot of the spectral electric energy $|E_{k}|^{2}$ , evaluated at $t=2000$ (in black), at $t=4500$ (in red), at $t=7000$ (in green) and at $t=32000$ (in blue), as a function of the wave number $k$.}
    \label{fig:spettro}
\end{figure}

\begin{figure}[ht]
\centering
\includegraphics[width=0.49\textwidth]{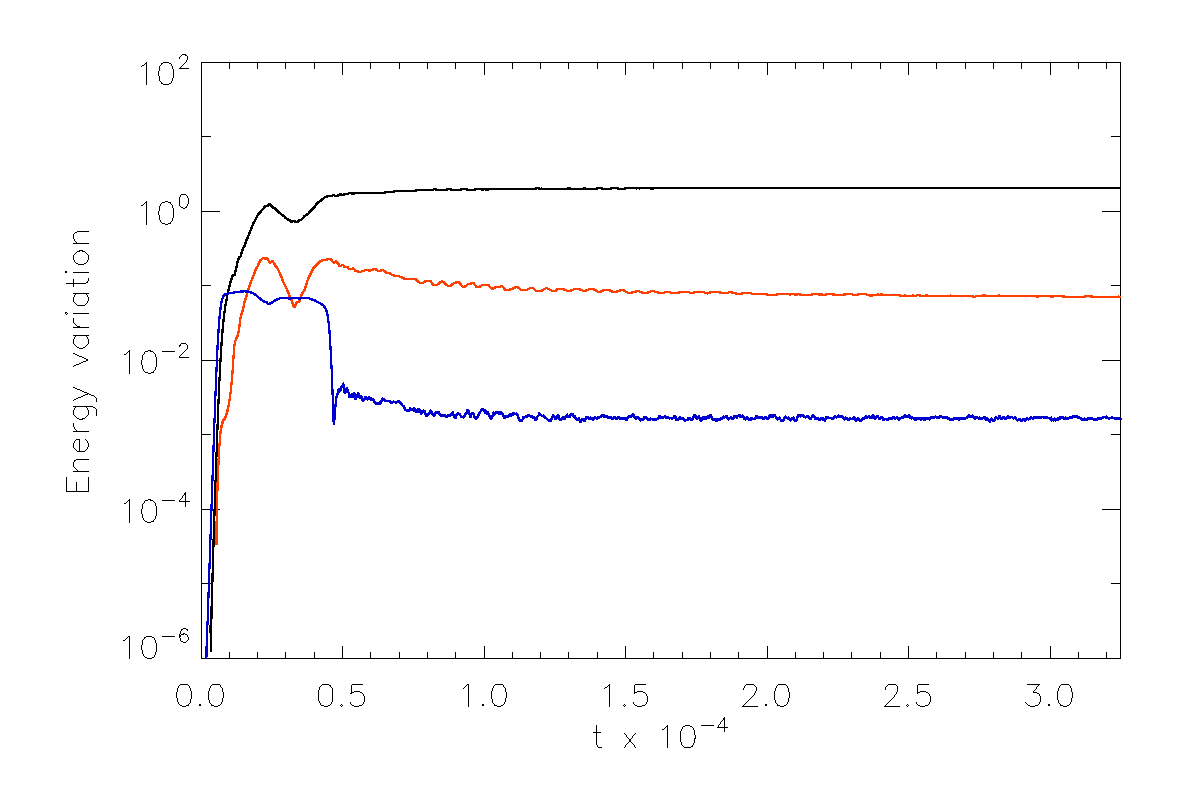} 
\caption{Time evolution of the variations in kinetic energy for protons (black curve) and electrons (red curve), along with the variation in electric energy (blue curve), presented on a semi-logarithmic scale.}
\label{fig:energy}
\end{figure}

\begin{figure*}[ht]
    \centering
    \includegraphics[width=0.24\textwidth]{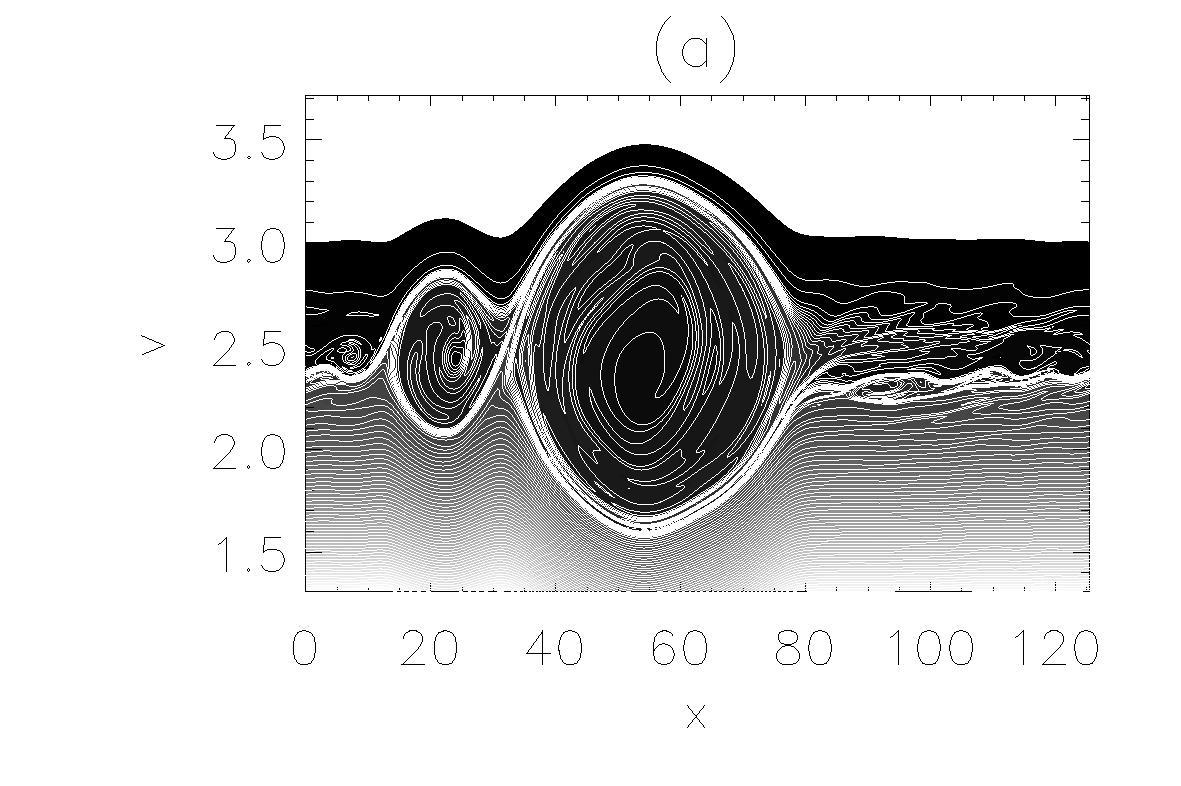}
    \includegraphics[width=0.24\textwidth]{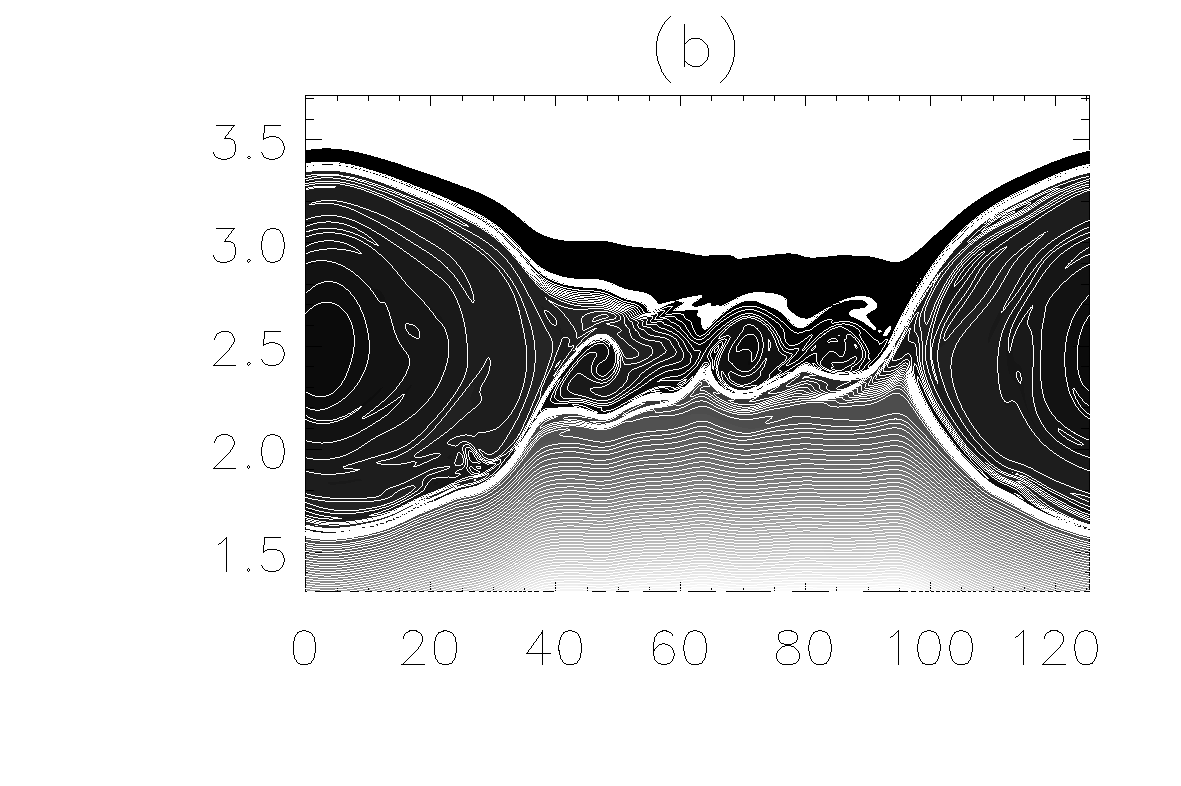}
    \includegraphics[width=0.24\textwidth]{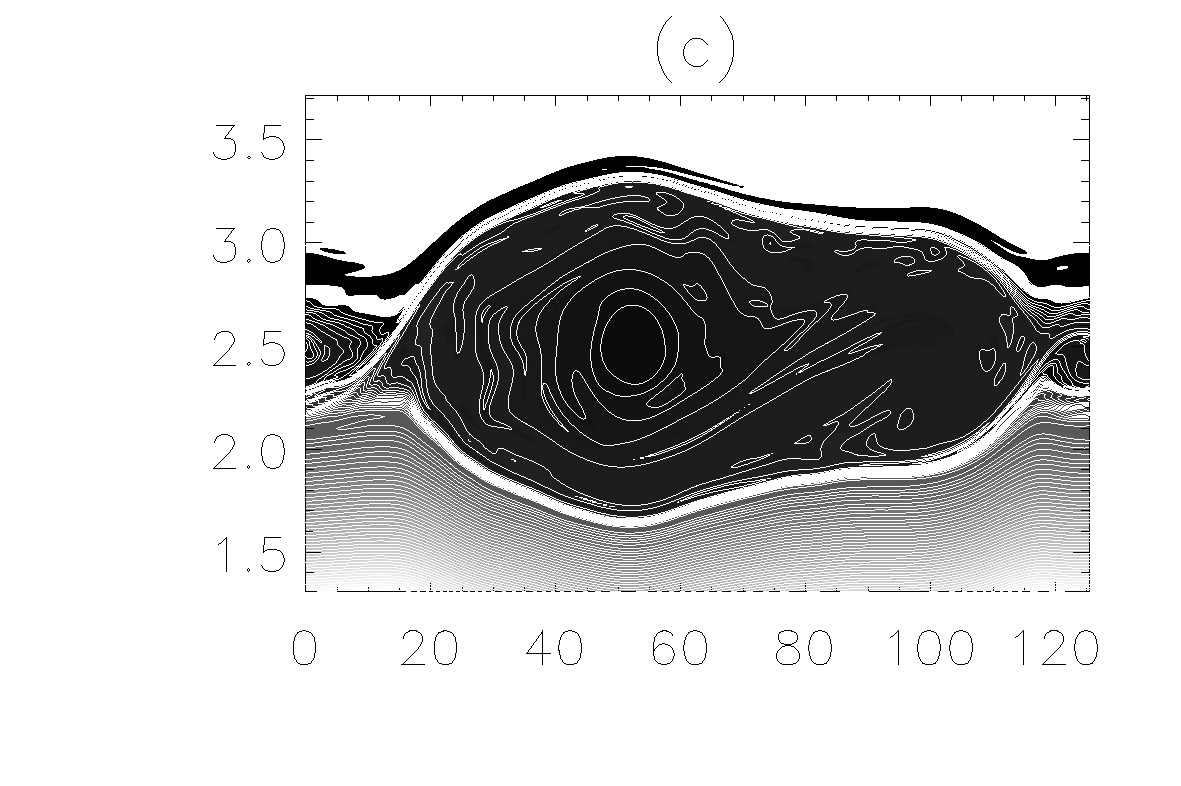}
    \includegraphics[width=0.24\textwidth]{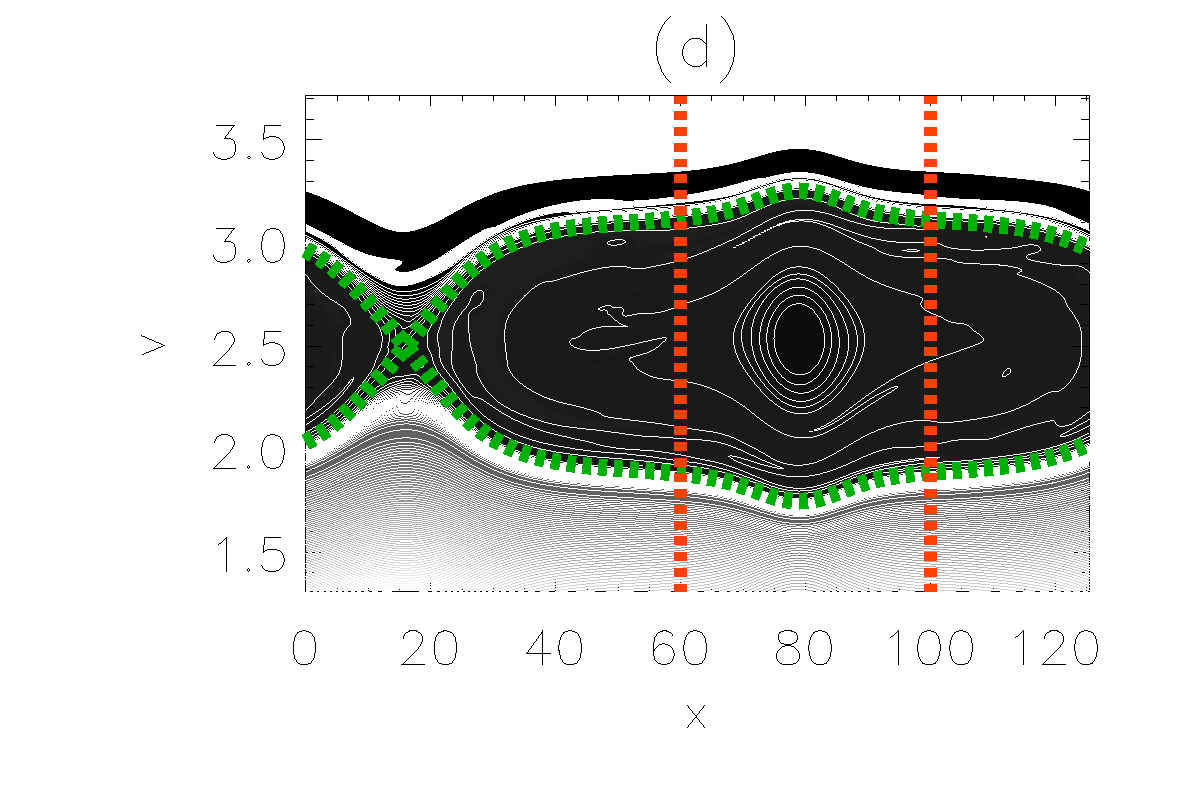}
    \includegraphics[width=0.24\textwidth]{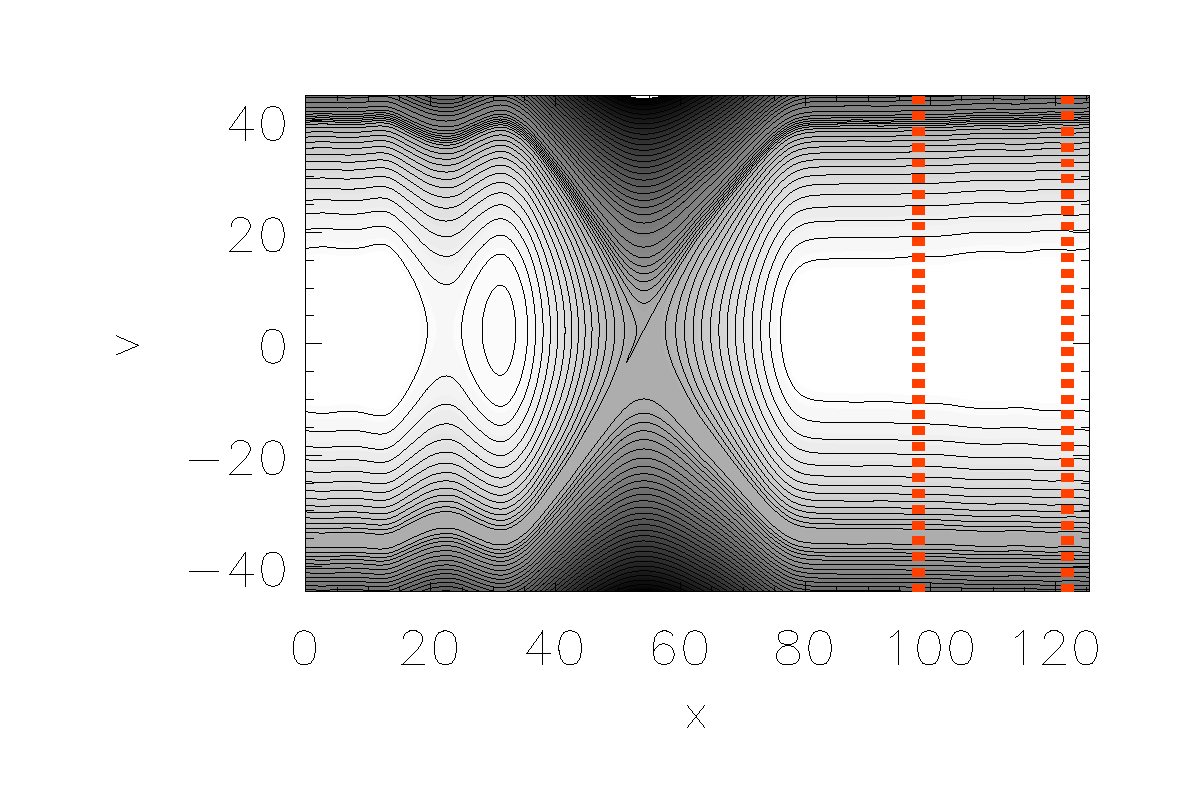}
    \includegraphics[width=0.24\textwidth]{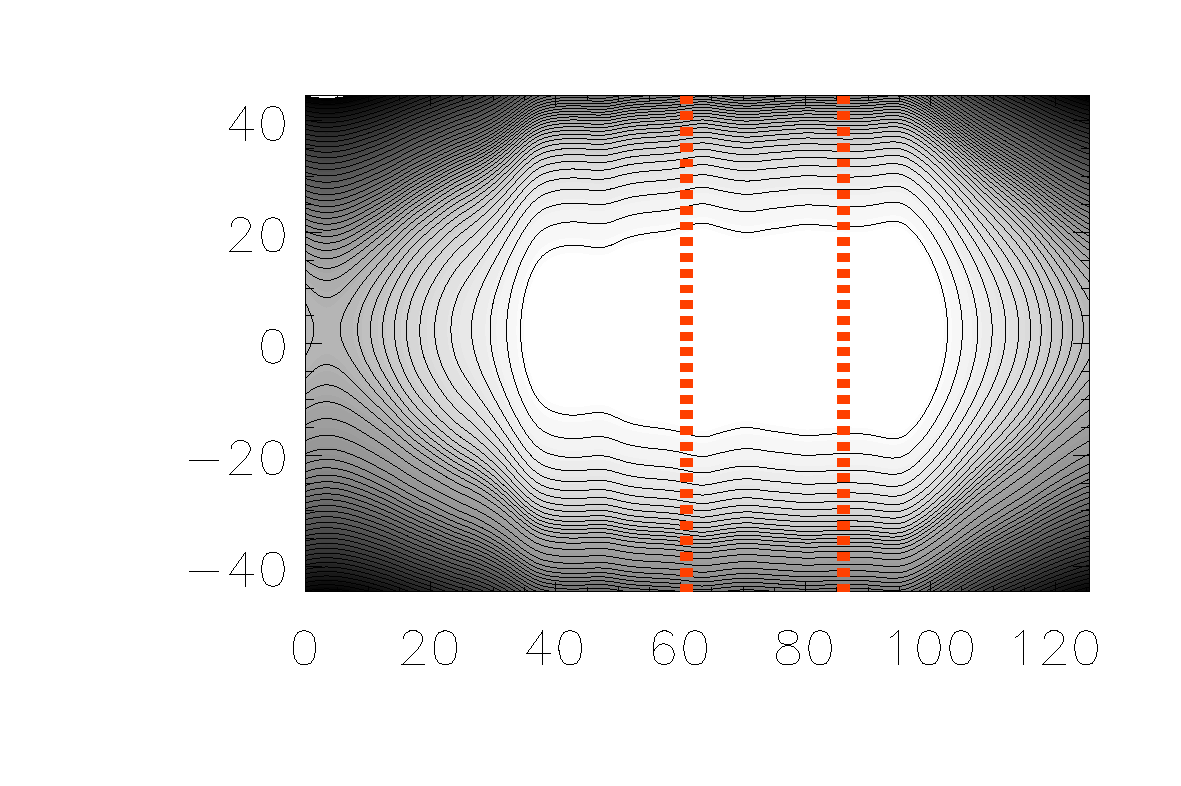}
    \includegraphics[width=0.24\textwidth]{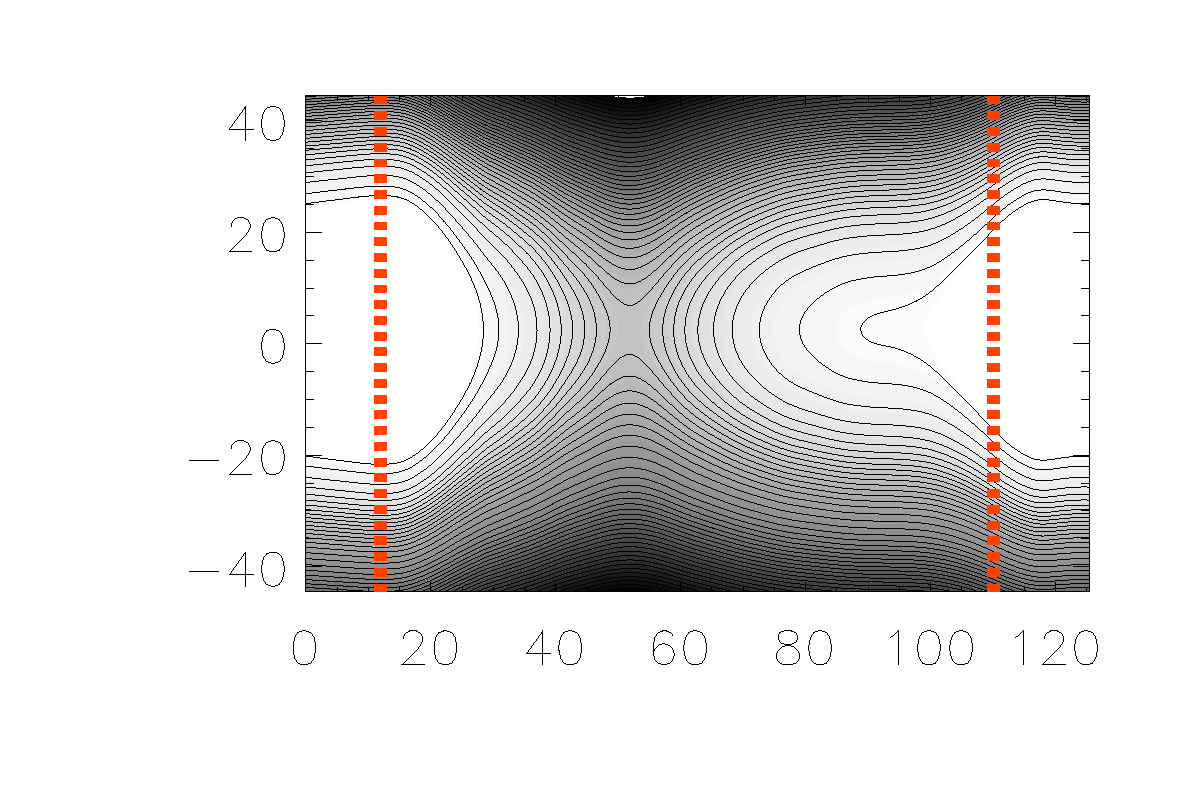}
    \includegraphics[width=0.24\textwidth]{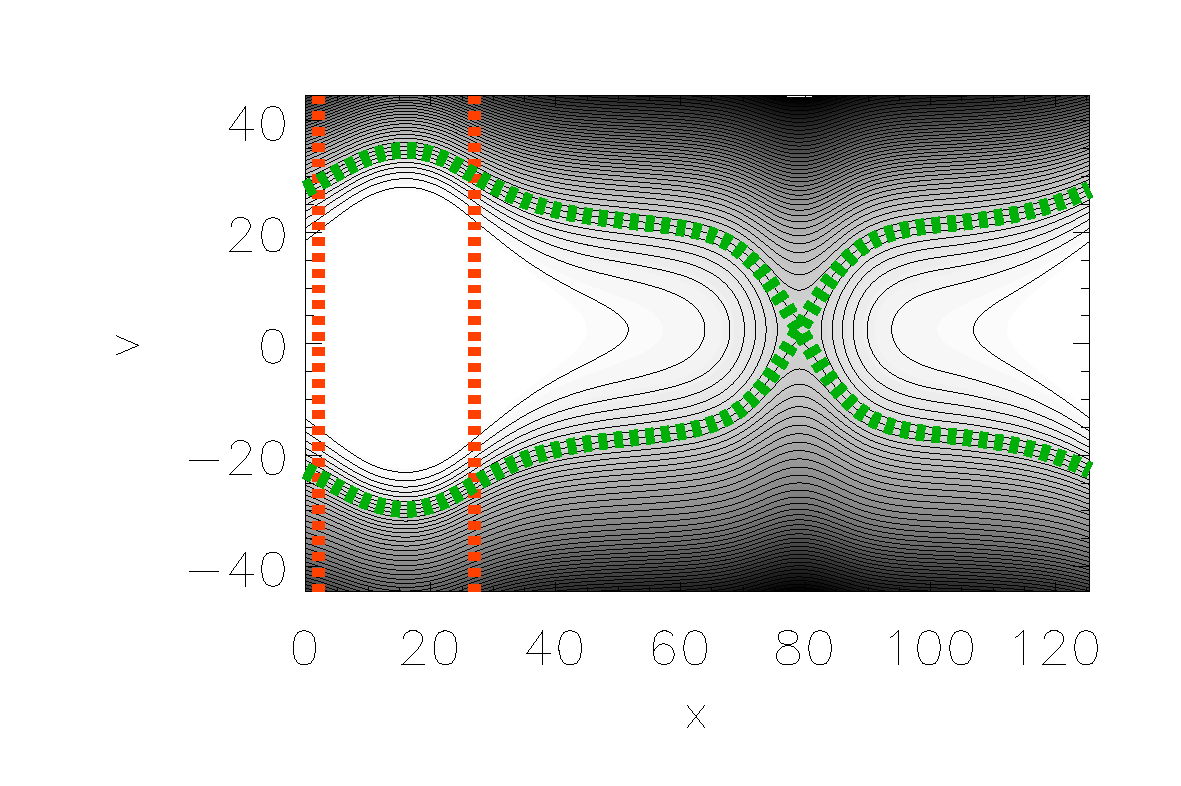}
   \caption{Top plot: Contour plots of the proton phase space distribution function at \( t=2000 \) (a), \( t=4500 \) (b), \( t=7000 \) (c), and \( t=32000 \) (d). Bottom plot: Contour plots of the electron phase space distribution function at the same time instances. The vertical red-dashed lines in panels (d) indicate the interval used for calculating the spatial averages shown in Fig. \ref{fig:fd_vs_v}. The green-dashed lines in panels (d) represent the theoretical expressions of the separatrices, $v_{p,\pm} = v_{\phi} \pm \sqrt{2\left(\phi_{max} - \phi\right)}$ for protons and $v_{e,\pm} = v_{\phi} \pm \sqrt{2\frac{m_{p}}{m_{e}}\left(-\phi_{max} + \phi\right)}$ for electrons, where $\phi$ is the electric potential and $\phi_{max}$ its maximum value.}
    \label{fig:contour}
\end{figure*}

\begin{figure*}[ht]
    \centering
  \includegraphics[width=0.49\textwidth]{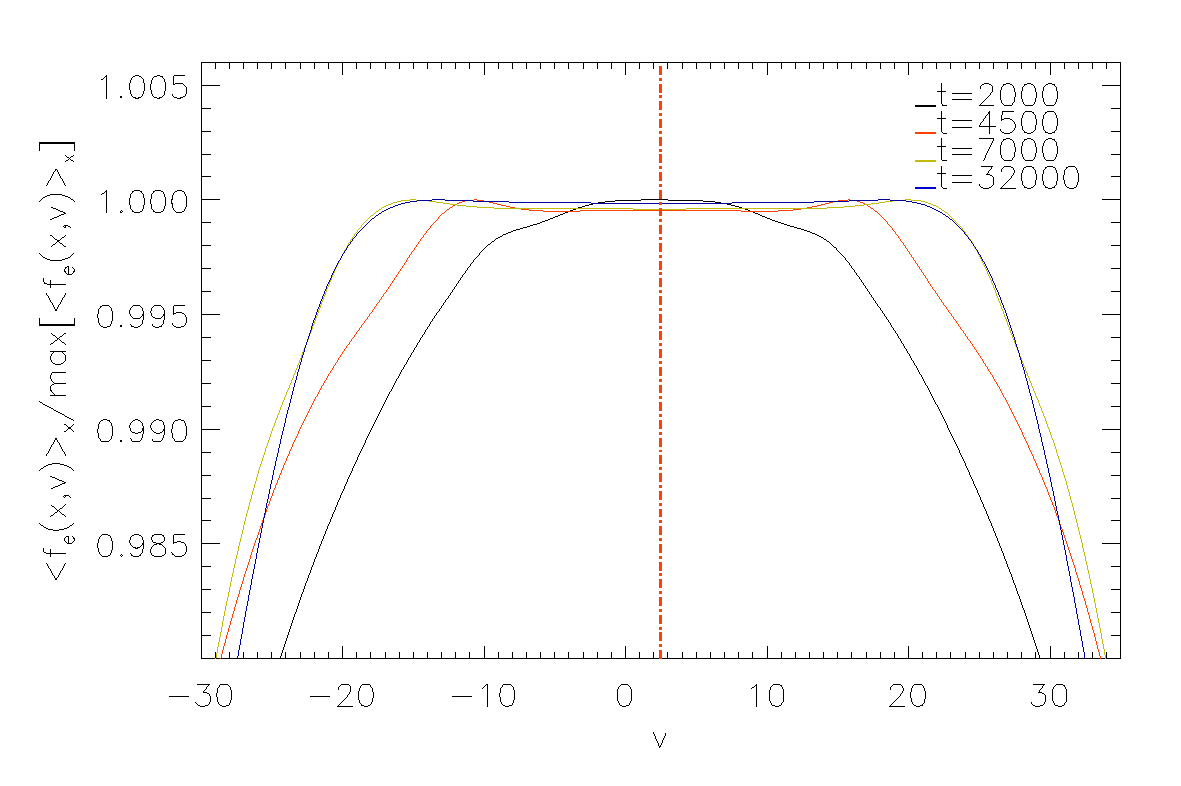} 
    \caption{One-dimensional spatial averages of the electron distribution functions, calculated over the intervals defined by the red-dashed lines in the bottom plots of Fig.~\ref{fig:contour}. The distributions are normalized to their respective maximum values and shown at successive times in the simulation, represented by the black, red, yellow, and blue curves, respectively. The vertical red dot-dashed line represents the phase speed ($v_\phi\simeq 2.5$) of the IBk fluctuations.}
    \label{fig:fde_all}
\end{figure*}

\begin{figure*}[ht]
    \centering
  \includegraphics[width=0.49\textwidth]{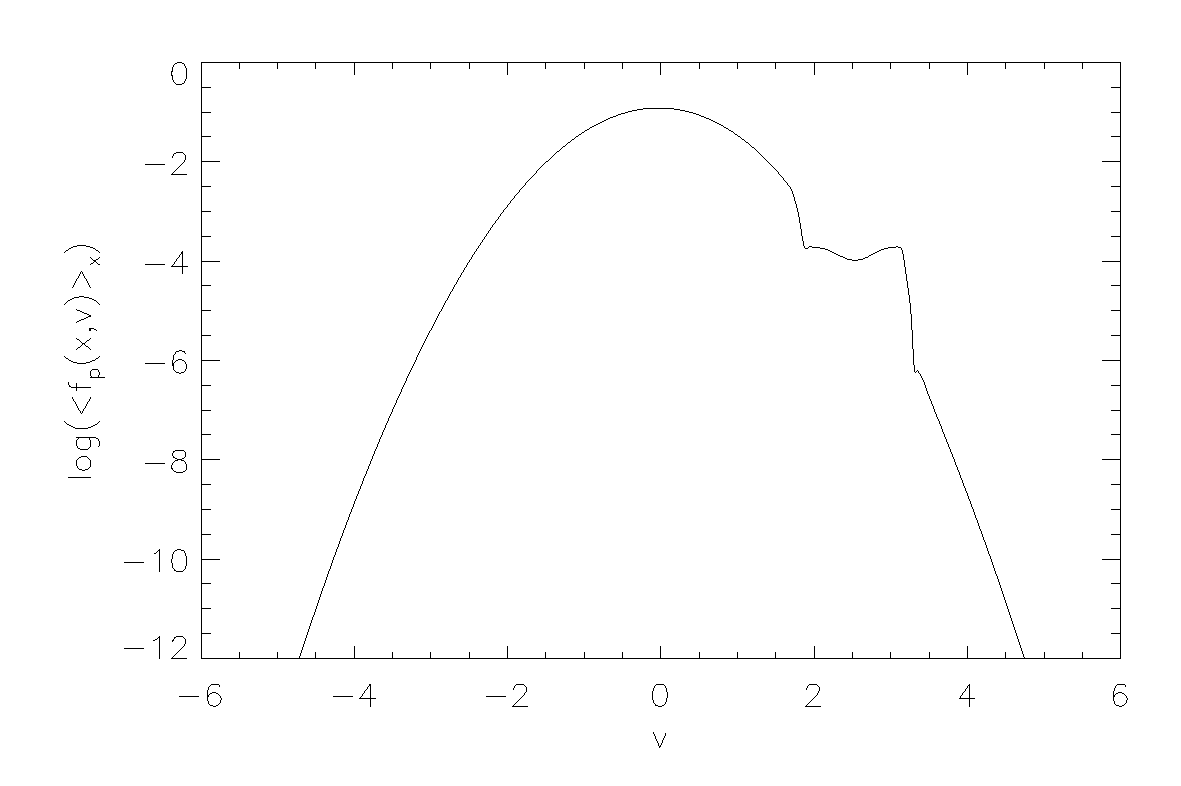} \hfill
   \includegraphics[width=0.49\textwidth]{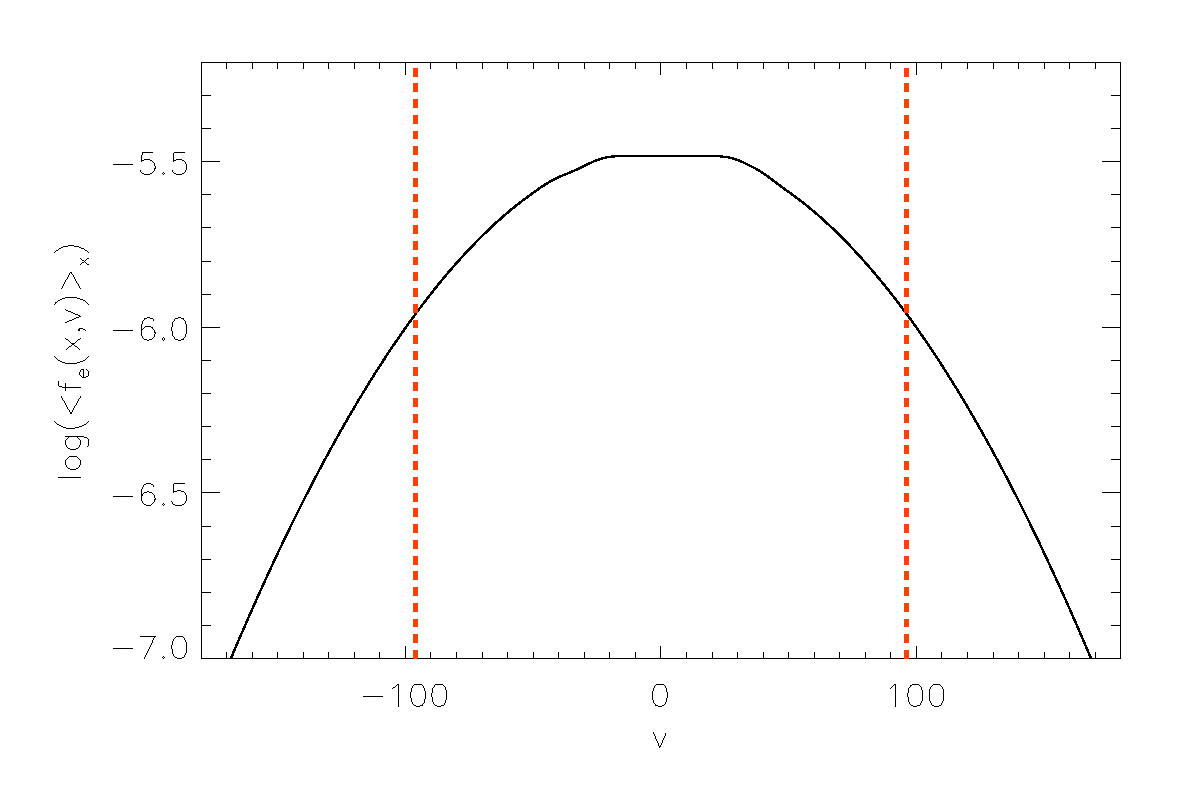} 
    \caption{1D spatial average of the proton distribution function (left plot), computed over the range \( x \in [60, 100] \), and the electron distribution function (right plot), computed over the range \( x \in [2, 27] \). The vertical red-dashed lines in the right plot indicate the electron thermal speed \( v_{th, e} \).}
    \label{fig:fd_vs_v}
\end{figure*}

\begin{figure*}[ht]
    \centering
  \includegraphics[width=0.40\textwidth]{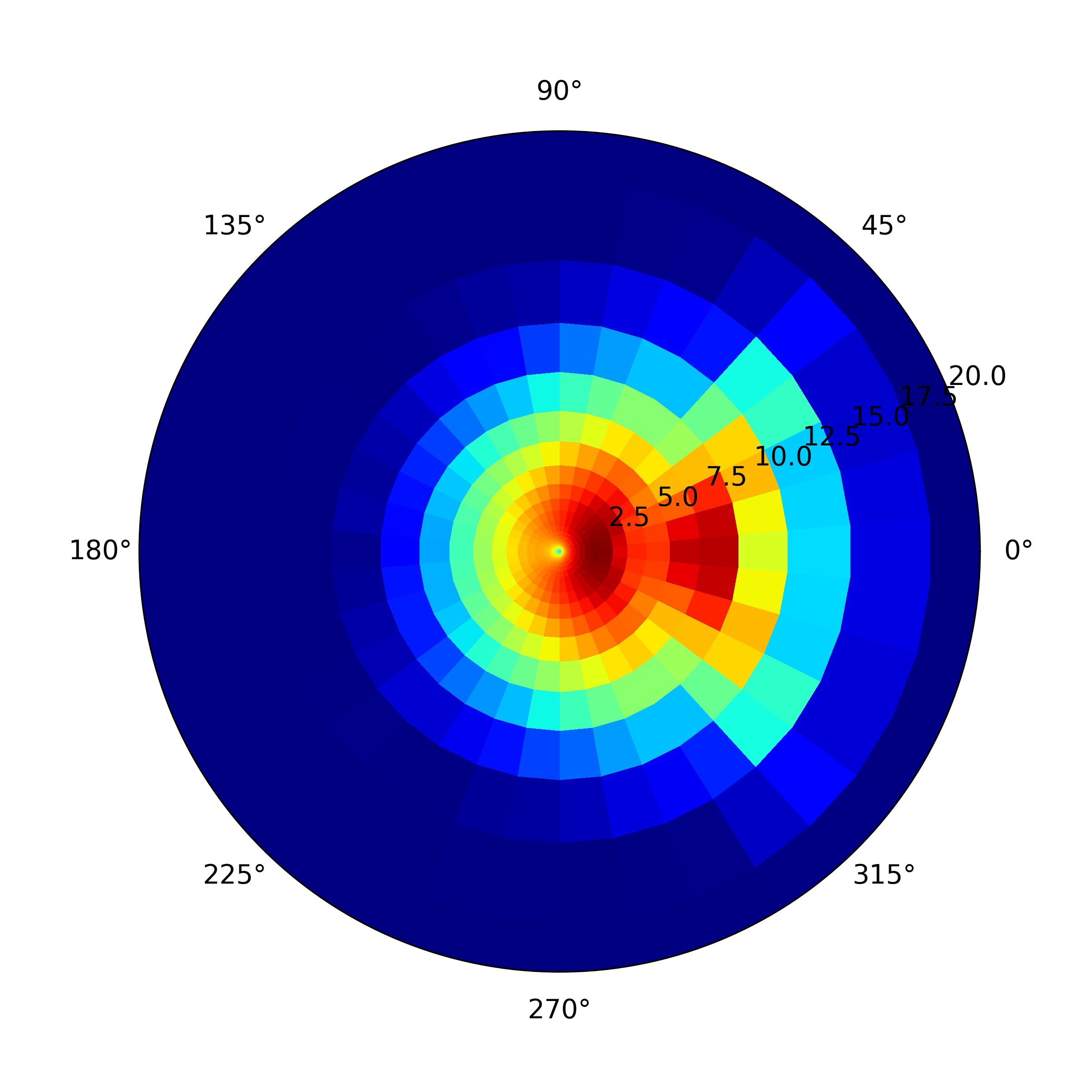} \hfill
   \includegraphics[width=0.40\textwidth]{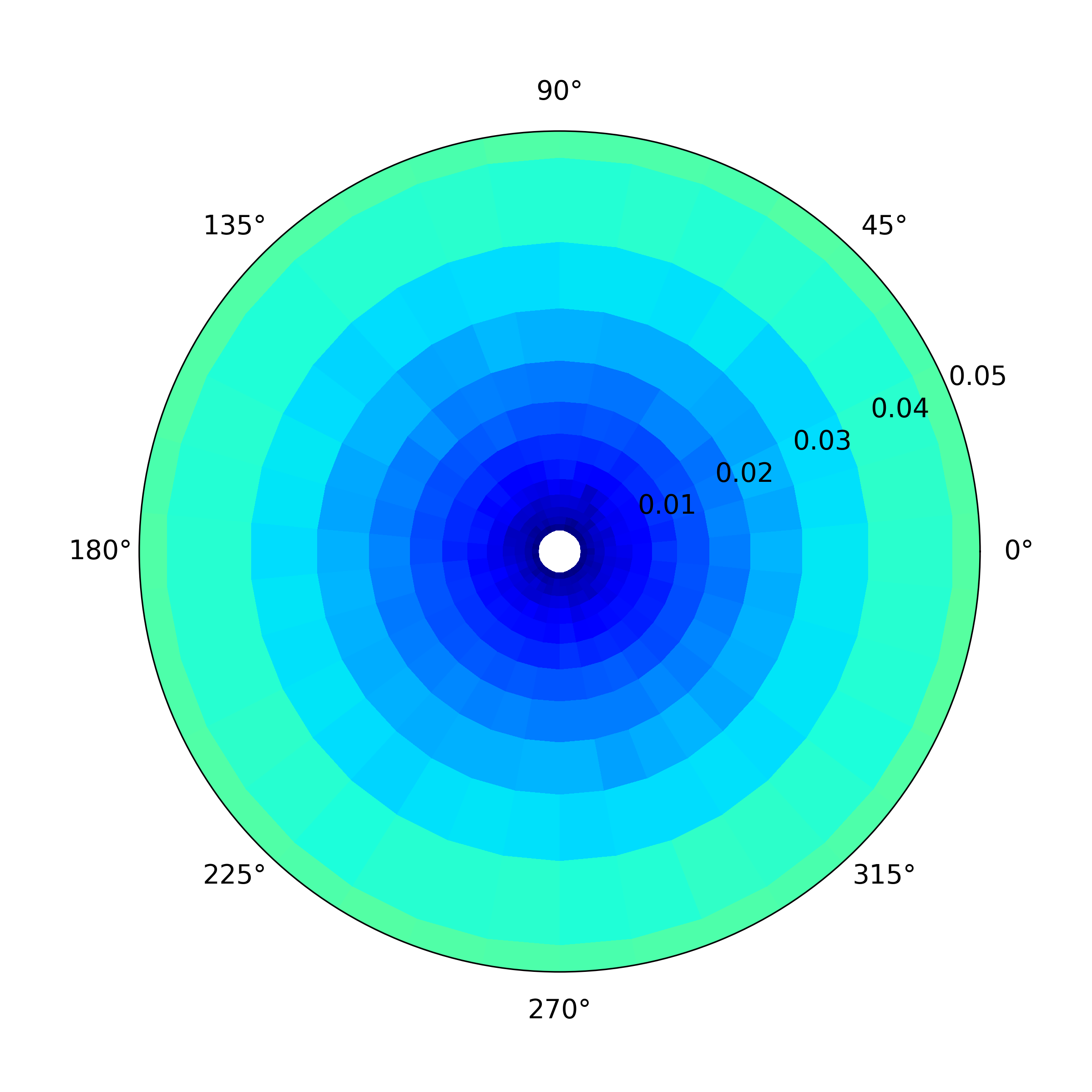} 
   \caption{Virtual electrostatic analyzer (top-hat) measurements of the proton (left plot) and electron (right plot) velocity distributions from the simulation. Both plots are shown in the \(\left(E, \phi\right)\) plane for \(\theta = \pi/2\).}
    \label{fig:Top_hat}
\end{figure*}

\end{document}